\documentclass[pra,twocolumn,showpacs,floatfix,a4paper,superscriptaddress]{revtex4}
\usepackage{bm,color,graphicx,amsmath,txfonts}
\usepackage{hyperref}

\newcommand{\rp}[1]{(\ref{#1})}

\newcommand{\abs}[1]{\left|{#1}\right|}

\newcommand{\av}[1]{\left\langle #1 \right\rangle}

\newcommand{\da}{^\dagger}

\newcommand{\pt}[1]{\left( #1 \right)}
\newcommand{\pq}[1]{\left[ #1 \right]}

\newcommand{\ii}{{\rm i}}

\newcommand{\nn}{{\nonumber}}

\newcommand{\FF}{{\cal F}}

\newcommand{\NN}{{\cal N}}

\begin{document}

\title{Discriminating the effects of collapse models from environmental diffusion \\ with levitated nanospheres}

\author{Jie Li}
\affiliation{School of Science and Technology, Physics Division, University of Camerino, I-62032 Camerino (MC), Italy}
\author{Stefano Zippilli}
\affiliation{School of Science and Technology, Physics Division, University of Camerino, I-62032 Camerino (MC), Italy}
\affiliation{INFN, Sezione di Perugia, I-06123 Perugia, Italy}
\author{Jing Zhang}
\affiliation{State Key Laboratory of Quantum Optics and Quantum Optics Devices, Institute of Opto-Electronics, Shanxi University, Taiyuan 030006, China}
\author{David Vitali}
\affiliation{School of Science and Technology, Physics Division, University of Camerino, I-62032 Camerino (MC), Italy}
\affiliation{INFN, Sezione di Perugia, I-06123 Perugia, Italy}

\begin{abstract}
Collapse models postulate the existence of intrinsic noise which modifies quantum mechanics and is responsible for the emergence of macroscopic classicality. Assessing the validity of these models is extremely challenging because
%, although their expected effects can be significant in various realistic situations,
it is non-trivial to discriminate unambiguously their presence in experiments where other hardly controllable sources of noise compete to the overall decoherence. Here we provide a simple procedure able to probe the hypothetical presence of the collapse noise with a levitated nanosphere in a Fabry-P\'erot cavity. We show that the stationary state of the system is particularly sensitive, under specific experimental conditions, to the interplay between the trapping frequency, the cavity size, and the momentum diffusion induced by the collapse models, allowing to detect them even in the presence of standard environmental noises.
\end{abstract}


\date{\today}
\maketitle

\section{INTRODUCTION}

Quantum mechanics (QM) has been proven as an extremely accurate theory for describing objects at microscopic scales. It intrinsically shows no limit to describe large massive systems~\cite{Leggett,Hammerer,Sciarrino,Arndt}. However, the lack of observations of macroscopically distinguishable superposition states of macroscopic objects allows to conjecture that QM should be modified at large scales. The modifications are aimed to explain the collapse of the wave function at the macroscopic level and also to solve the quantum measurement problem~\cite{Bassi}. Various attempts have been made in this direction, including the Ghirardi-Rimini-Weber approach~\cite{GRW}, continuous spontaneous localization (CSL)~\cite{CSL}, and gravitationally-induced collapse models (CMs)~\cite{DP}. These models modify the Schr\"odinger equation by introducing appropriate stochastic non-linear terms, which resolve the problems at macroscopic scales, while reproducing the standard results at microscopic scales~\cite{BassiRMP}.

Whether the proposed CMs are or are not exact should be examined by experiments. Matter-wave interferometry~\cite{ArndtRMP}, where large massive molecules or clusters~\cite{cluster} are sent through interference gratings, is considered as one of the ideal arenas to test CMs. However, the mass range has, so far, yet to be reached to effectively test CMs. An alternative approach is based on cavity optomechanics~\cite{OMRMP}, where one could prepare massive mechanical resonators (MRs) in linear superposition states and monitor their decoherence~\cite{mirror,nanosphere,Ulbricht}. More recent proposals~\cite{Mauro,Nimmrichter,Sekatski,Diosi,Barker2} suggested to test CMs in optomechanical systems in a non-interferometric way, so that the preparation of large spatial superposition states is not required. In fact, the spontaneous collapse mechanism leads to spatial decoherence, i.e., momentum diffusion, of the MR, which results in additional phase noise of the light leaking out of the cavity~\cite{Mauro,Nimmrichter,Sekatski}.

In general, the identification of systems and regimes in which collapse-induced diffusion is theoretically dominant over the environmental noises is not sufficient for the design of experiments able to univocally decide whether a given observation is actually the result of collapse theories, or of other uncontrolled sources of environmental decoherence. Being able to differentiate unambiguously  their effect is a non-trivial task which deserves further studies and the identification of dedicated experimental procedures. The fundamental idea at the base of this work, is the observation that different sources of noise exhibit different scalings with the system parameters, and hence distinguishable scalings of measurable quantities are expected to be observed if determined sources of noise are or are not actually present or effective. Specifically, in this article, we discuss a novel and efficient test of the CSL model in optomechanical systems in the regime of high mechanical quality factors and cryogenic temperatures, that is realizable with trapped levitated nanospheres in Fabry-P\'erot optical cavities~\cite{Chang,nanosphere,Barker,Raizen,gieseler,Markus}.
We demonstrate that, in these systems, the different noise sources are particularly sensitive to the trapping frequency $\omega$ and to the length of the cavity $L$, so that the validity of the CSL model can be actually probed by the study of the nanoparticle dynamics as a function of $\omega$ and $L$. This observation is general and can be applied to any optomechanical scheme. In the following, we explore its effectiveness in the analysis of the system steady state,
by analyzing the phase noise of the light field leaking out of the cavity as a function of $\omega$ and $L$. We show that, in experimentally achievable parameter regimes, a nanosphere can be prepared in a stationary state particularly sensitive to the mechanical momentum diffusion induced by the CMs, and, most importantly, whose statistical properties scale differently depending on whether CMs are true or not, hence allowing for an efficient test of CMs.

The remainder of this paper is organized as follows: In Sec.~\ref{model} we describe in detail our model, provide the Langevin equations governing the dynamics, and analyze various conventional diffusion rates of the system as well as the nontrivial diffusion rate induced by the collapse noise postulated in the CSL model. In Sec.~\ref{AAA} and~\ref{BBB}, we provide two alternative proposals to effectively test the CSL model, i.e., by varying the trapping frequency and the cavity length. We show details on the pressure and temperature, as well as the degree of precision for the measurement, required for testing the corresponding value of the collapse rate. Finally, in Sec.~\ref{concl} we draw our conclusions and in the Appendix we discuss the effects of blackbody radiation in our system.

\section{The model}
\label{model}

We consider a single nanosphere of radius $R$ trapped by a harmonic dipole trap, at frequency $\omega$, within a Fabry-P\'erot optical cavity with length $L$, finesse $\FF$, and with mirror radius of curvature $R_c$. A single cavity mode, with resonance frequency $\omega_c$, is driven by an external field at power $P$ and frequency $\omega_L$, detuned by $\Delta=\omega_c-\omega_L$, and is coupled to the center of mass of the  nanosphere~\cite{Chang,nanosphere,Barker,Raizen,gieseler,Markus}.
The relevant degrees of freedom for the linearized system dynamics are
the fluctuations of the cavity field and of the mechanical center of mass variables about their respective average values, described by the bosonic operators $a$ and $a\da$ (with $[a,a\da]=1$) for the cavity field, and by the dimensionless position and momentum $x$ and $p$ (with $[x,p]=\ii$) for the nanoparticle. The corresponding quantum Langevin equations (QLEs) read
\begin{eqnarray}\label{QLEnum}
\dot a&=&-\pt{\ii\,\Delta+\kappa}\, a-\ii\, G\, x+\sqrt{2\kappa}\,a^{\rm in},
\\
\dot x&=&\omega\, p,
\nn\\
\dot p&=&-\omega\, x -\gamma\, p-G\pt{a+a\da}+F_{\rm air}+F_D,
\nn
\end{eqnarray}
where the linearized coupling strength $G=g\,\alpha$ is proportional to the average cavity field $\alpha= \sqrt{2\kappa P/\pq{\hbar\omega_L\pt{\Delta^2+\kappa^2}}}$, with $g$ the bare optomechanical coupling, which can be expressed as $g{=}\omega_c\sqrt{\frac{\hbar}{m\omega}}\frac{2\pi}{\lambda_c}\frac{\epsilon-1}{\epsilon+2}\frac{3V_s}{4V_c}$~\cite{Chang}, with $\lambda_c$ the cavity wavelength, $\epsilon$ the electric permittivity of the nanosphere, $V_s$ its volume, and $V_c{=}\pi L W_0^2/4$ the cavity mode volume with mode waist $W_0=[\lambda_c L (2R_c/L{-}1)^{1/2}/2\pi]^{1/2}$. $\kappa{=} \pi c/(2\FF L)$ is the cavity linewidth with $c$ the speed of light, and $\gamma$ is the damping rate of the mechanical motion. For levitated nanospheres $\gamma$ can be extremely small, resulting in very high quality factors, $\gtrsim10^{10}$~\cite{Oriol2011},
with the dominant contribution due to friction from residual air molecules, for which
$\gamma=\frac{16}{\pi}\frac{P_a}{\bar v\,R\,\rho_0}$, with $P_a$ the gas pressure, $\bar v=\sqrt{3k_B T/m_a}$ the mean speed of the air molecules, $m_a$ their mass (which we take $m_a=28.97$ amu), and $T$ the air temperature~\cite{Chang}.
In Eq.~\rp{QLEnum} we have included the relevant sources of noise affecting the dynamics of the system, and leading to mechanical Brownian motion, in terms of the $\delta$-correlated stochastic forces $a^{\rm in}(t)$, $F_{\rm air}(t)$ and $F_D(t)$ (see the Appendix for a comment on the effects of blackbody radiation). Firstly, $a^{\rm in}(t)$ is the input noise operator for the cavity field due to the fluctuations of the external electromagnetic environment. Its only non-zero correlation function is
\begin{equation}
 \langle a^{\rm in}(t)\,{a^{\rm in}}\da(t')\rangle=\delta(t-t'). 
\end{equation}
The term $F_{\rm air}(t)$, instead, accounts for the mechanical noise due to the scattering of background air molecules which is related to the dissipation rate $\gamma$ by the fluctuation-dissipation theorem, and its autocorrelation function, in the relevant high temperature regime, is given by 
\begin{equation}
\av{F_{\rm air}(t)\,F_{\rm air}(t')}=D_a\delta(t-t') 
\end{equation}
with the diffusion rate $D_a=2\gamma\, k_B T/(\hbar\omega)$. Finally, $F_D(t)$ accounts for pure diffusion of the particle motion with autocorrelation function
\begin{equation}
\av{F_D(t)\,F_D(t')}=(D+\lambda_{\rm sph})\delta(t-t'), 
\end{equation}
responsible for dephasing and decoherence on the nanosphere. Here we have separated the contributions $D$ and $\lambda_{\rm sph}$ which describe, respectively, the effects of light scattering and of collapse-induced diffusion.

In the framework of the CSL model, the collapse-induced diffusion rate for a spherical nanoparticle, with constant mass density $\rho_0$ and harmonically trapped at frequency $\omega$, is given by~\cite{Nimmrichter}
\begin{equation}
\lambda_{\rm sph}=\frac{\hbar}{\omega}\frac{8\pi\,\lambda\,\rho_0}{m_0^2}\left[e^{-R^2/r_c^2}-1+\frac{R^2}{2r_c^2}(e^{-R^2/r_c^2}+1)\right]\frac{r_c^4}{R^3} \\ ,
\label{lambda_sph}
\end{equation}
with $m_0$ the atomic mass unit.
%which achieves a maximum around $R\approx2.38\,r_c$ at which the collapse effect is most prominent.
The actual strength of collapse noise is determined by two phenomenological parameters,
the characteristic length $r_c$ and the collapse rate $\lambda$. While there is significant agreement on the estimated value for the characteristic length $r_c\simeq 100$ nm, the expected value of $\lambda$ is more controversial. The initial estimate of $\lambda$ is $~10^{-16}$ s$^{-1}$~\cite{GRW,CSL}, however, larger values have been proposed by other authors (for example  $~10^{-8\pm2}$ s$^{-1}$ in Ref.~\cite{Adler}). In any case, different experimental results provide indications that  $\lambda$ should be lower than $~10^{-8}$ s$^{-1}$~\cite{Vinante}, $~10^{-9}$ s$^{-1}$~\cite{Pearle2014} and $~10^{-11}$ s$^{-1}$~\cite{Curceanu}, for $r_c\simeq 100$ nm.
Relevant new proposals should be able to confirm or improve such results by either lowering the upper bound of $\lambda$ or by detecting noise effects which cannot be explained by standard decoherence.

The diffusion rate via photon scattering, instead, can be expressed as the sum of the contributions due to the scattering of trapping and cavity light, i.e., $D{=}D_t{+}D_c$, which are given, respectively, by~\cite{Pflanzer}
\begin{equation}
D_t=\frac{8\epsilon_c^2 k_c^6 R^3}{9 \rho_0 \omega}\frac{{\cal I}}{\omega_{Lt}}\ ,
\hspace{1.1cm}
D_c=\frac{2\epsilon_c^2 k_c^6 R^3}{9 \rho_0 \omega}\frac{\hbar n_{\rm ph}c}{V_c}\ ,
\label{diff}
\end{equation}
where $\epsilon_c=3\frac{\epsilon-1}{\epsilon+2}$, $k_c=2\pi/\lambda_c$, $\omega_{Lt}$ is the frequency of the trapping laser, and the trapping frequency is determined by $\omega=[4\epsilon_c {\cal I}/(\rho_0 c W_t^{2})]^{1/2}$, with $W_t$ the waist of trapping light which can be approximated by $W_t\approx\lambda_c/(\pi{\cal N})$ with ${\cal N}$ the numerical aperture, and ${\cal I}$ the intensity of the trapping field, which is related to the power by ${\cal I}=P_t/(\pi W_t^{2})$. Finally $n_{\rm ph}=\abs{\alpha}^2$ is the mean cavity photon number.

We observe that $D_t$ and $D_c$ increase while $D_a$ decreases with the size of the nanoparticle. We have checked that the optimal size of the sphere for testing the CSL theory is roughly around $r_c$~\cite{Nimmrichter}. We also note that large field powers imply large diffusion rates by light scattering and as a result low powers are in general required to reduce the photon scattering induced diffusion and to make it comparable to the collapse noise, entering therefore a regime in which CSL could be detected. These observations are however not sufficient to discriminate the effect of collapse induced diffusion. In order to achieve this, we have to identify strategies which exhibit a qualitatively different response when CSL is present and when it is not. In this respect, the central observation of this work is that the diffusion rates exhibit very peculiar scalings as a function of $\omega$ and $L$. As we will discuss below different scalings of the diffusion rates imply distinguishable steady state behavior of the nanosphere which may be exploited to distinguish the effects of CMs. Specifically, in our setup, where we use a weak driving field at a frequency smaller than the cavity frequency which yields a weak cooling force on the mechanical motion and stabilizes it, the position fluctuation gets encoded in the phase quadrature of the cavity field $Y=-\ii(a-a\da)/\sqrt{2}$, such that $\langle x^2\rangle \propto \langle Y^2 \rangle+\text{shot noise}$. Consequently direct information on the nanosphere diffusion can be extracted by the measurement of the optical phase at the cavity output. Hereafter we analyze the steady state behavior of $\langle Y^2 \rangle$ in various experimental conditions, versus either $\omega$ or $L$, which may be used to probe the strength of $\lambda$.

\subsection{Test of the CSL model by varying the trapping frequency}
\label{AAA}

\begin{figure}[t!]
\includegraphics[width=1\linewidth]{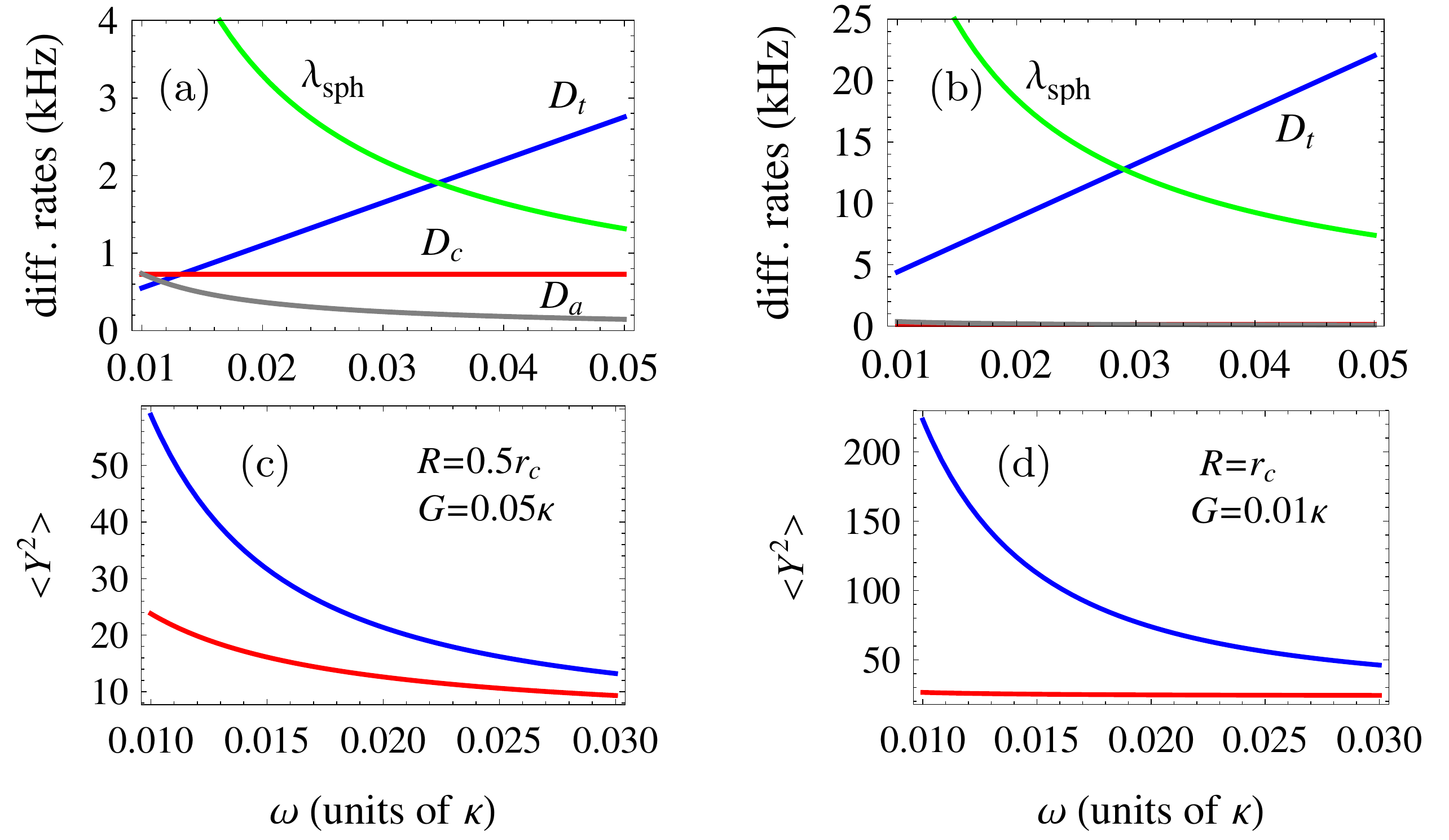}
\caption{(a), (b): Diffusion rates for the scattering of trapping light $D_t$ (blue), cavity light $D_c$ (red) and air molecules $D_a$ (gray), and the collapse rate $\lambda_{\rm sph}$ (green) versus the mechanical frequency $\omega$ with (a) $R=0.5r_c$ and $G=0.05\kappa$; (b) $R=r_c$ and $G=0.01\kappa$. Note that the curves of $D_c$ and $D_a$ in (b) are very close to the $\omega$-axis and no longer visible. (c), (d): Steady state variance of the optical phase quadrature $\langle Y^2\rangle$ versus the trapping frequency $\omega$. Blue (red) lines refer to the case with (without) the CSL effect. The parameters for (c)/(d) correspond to those for (a)/(b). The other parameters are $L=R_c=1$ cm, $\FF=10^5$ (corresponding to $\kappa=0.47$ MHz), $\Delta=0.01\kappa$, $\lambda_c=1064$ nm, $\NN=0.6$, $T=1$ K, $P_a=10^{-10}$ Torr, $\lambda=10^{-8}$ s$^{-1}$, $r_c=100$ nm, and we consider a diamond nanosphere with $\rho_0=3.5$ g/cm$^3$ and $\epsilon=5.76$.}
\label{variomega}
\end{figure}

\begin{figure}[b]
\includegraphics[width=1\linewidth]{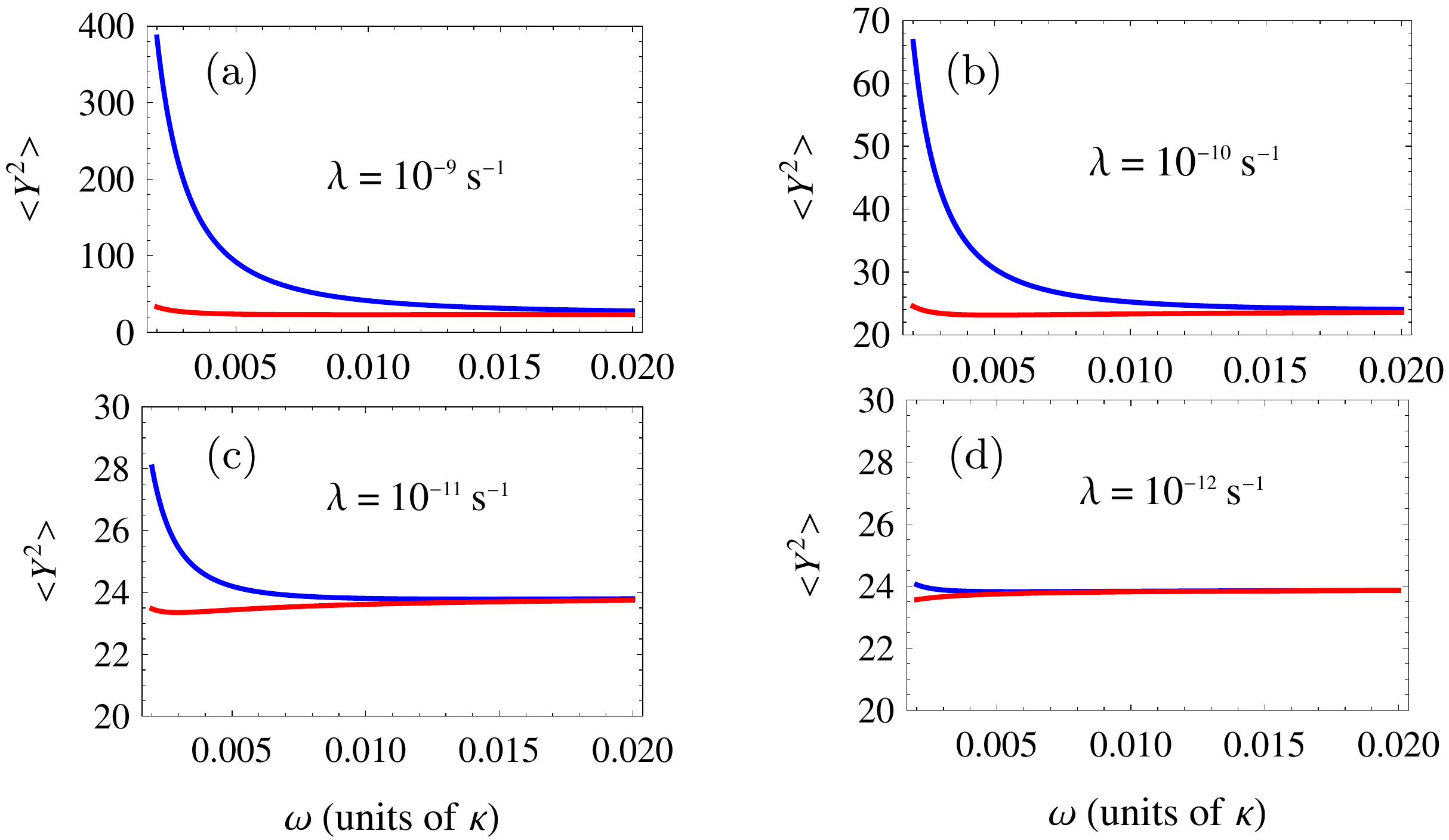}
\caption{Steady state variance of the optical phase quadrature $\langle Y^2\rangle$ versus the trapping frequency $\omega$ for (a) $\lambda=10^{-9}$ s$^{-1}$, $T=200$ mK, $P_a=10^{-10}$ Torr; (b) $\lambda=10^{-10}$ s$^{-1}$, $T=100$ mK, $P_a=3\times10^{-11}$ Torr; (c) $\lambda=10^{-11}$ s$^{-1}$, $T=60$ mK, $P_a=10^{-11}$ Torr; (d) $\lambda=10^{-12}$ s$^{-1}$, $T=10$ mK, $P_a=10^{-12}$ Torr. Blue (red) lines represent the case with (without) the CSL effect. In all plots we take $G=0.001\kappa$. The other parameters are as in Fig.~\ref{variomega} (b).}
\label{testlimit}
\end{figure}

We first note that $\lambda_{\rm sph}, D_a\propto 1/\omega$, $D_t\propto\omega$, and, when $G$ is fixed, $D_c$ is independent from $\omega$, as clearly shown in Fig.~\ref{variomega} (a). Specifically, $D_c$ can be made negligible for sufficiently small $G$, i.e., for sufficiently small driving power. Similarly, under the conditions of low pressure and temperature, $D_a$ is much smaller than $\lambda_{\rm sph}$ and $D_t$. This is the situation achieved in Fig.~\ref{variomega} (b), where we consider a larger size of the sphere $R=r_c$, in order to further reduce $D_a$ and increase $\lambda_{\rm sph}$. Thus, if in addition the power of the trapping light is small enough (corresponding to relatively small $\omega$), so that $D_t$ and $\lambda_{\rm sph}$ are of comparable strength, then the presence of spontaneous collapse mechanism can be demonstrated by detecting the output light as a function of the mechanical trapping frequency, which could be spanned by simply adjusting the intensity of the trapping light. In detail, the steady state variance of the optical phase quadrature displays distinguishably different behavior depending upon the presence or absence of collapses. This is shown in Fig.~\ref{variomega} (d) and Fig.~\ref{testlimit}, where we analyze the results for a diamond nanosphere with radius $R=100$ nm and a cavity with finesse $\FF=10^5$, and we compare the results  with (blue lines) and without (red lines) the effect of CSL.
In particular, the stationary variance $\langle Y^2 \rangle $ in the presence of the CSL effect increases rapidly as $\omega$ is gradually reduced, while it is practically independent upon $\omega$ without the CSL effect. Therefore, by repeating the experiment at different trapping light intensity one could verify this different behavior and determine the possible presence of spontaneous collapses.
When the value of $\lambda$ is reduced as in Fig.~\ref{testlimit}, the relative difference between the two curves (with and without the CSL) reduces. At $\lambda\sim10^{-12}$ s$^{-1}$, the two curves become hardly distinguishable, implying that the experimental realization of our protocol would allow, if no CSL effects are detected, to lower the upper bound of $\lambda$ to $10^{-12}$ s$^{-1}$.


\subsection{Test of the CSL model by changing the cavity length}
\label{BBB}

\begin{figure}[t]
\includegraphics[width=1\linewidth]{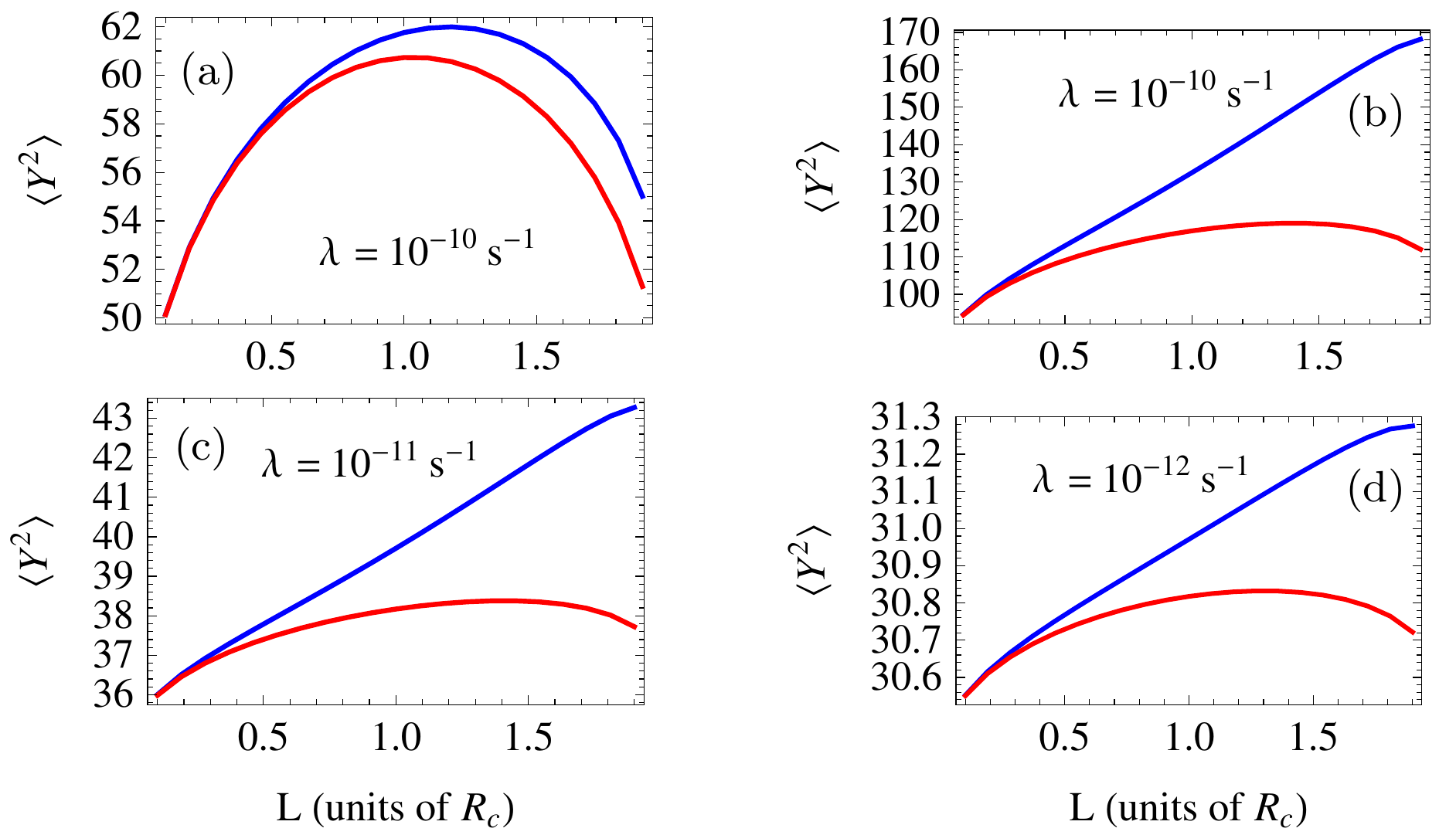}
\caption{Steady state variance of the optical phase quadrature $\langle Y^2\rangle$ versus the cavity length $L$ with (a) $\omega=\Delta=0.02\kappa$, $G=0.15\kappa$,  $P_a=10^{-10}$ Torr; (b) $\omega=\Delta=0.008\kappa$, $G=0.08\kappa$, $P_a=10^{-10}$ Torr; (c) $\omega=\Delta=0.008\kappa$, $G=0.025\kappa$, $P_a=10^{-11}$ Torr; (d) $\omega=\Delta=0.008\kappa$, $G=0.009\kappa$, $P_a=10^{-12}$ Torr.  The values of $\lambda$ are reported in each plot. All the curves are evaluated for $T=100$ mK and $R_c=2$ cm, and the other parameters are as in Fig.~\ref{testlimit}. Blue (red) lines represent the case with (without) the CSL effect. }
\label{figL}
\end{figure}

An alternative approach provides a more evident effect of collapse noise at the expense of a slightly more involved experimental protocol. In particular, we find that the diffusion rates show peculiar scalings with the size of the cavity $L$
%(at fixed radius of curvature $R_c$ of the mirrors)
when the ratios $\omega/\kappa$, $\Delta/\kappa$ and $G/\kappa$
are kept fixed.
Specifically, since $\kappa\propto 1/L$ then $\lambda_{\rm sph}\propto L$ and $D_{t}\propto 1/L$, while $D_{c}\propto \sqrt{2R_c/L-1}$ decreases with $L$ due to the particular scaling of the geometry of the cavity mode. In this case the experimental procedures should run as follows. For each value of the cavity length (implying different values of the cavity linewidth) the values of $\omega$, $G$ and $\Delta$ should be carefully tuned and monitored in order to achieve determined fixed values relative to $\kappa$, so that the system remains under the
same optomechanical condition. This can be achieved by varying the intensity and frequency of the stabilized driving and trapping laser, and simultaneously monitoring the $G$ and $\omega$ values. The resonance frequency $\omega$ is easily measured from the position of the resonance peak in the cavity output phase, while $G$ can be extracted from the nanosphere cooling rate, which is given by $G^2/\kappa$ in the bad cavity regime considered here. The steady state observables, detected for each set of parameters, will thus depend only on the diffusion rates. Thereby, if the powers of both trapping and driving light are sufficiently small (corresponding to relatively small $\omega$ and $G$), so that the $D_t$, $D_c$ and $\lambda_{\rm sph}$ are of comparable strength, then the behavior of the system steady state versus $L$ may distinguish the action of CMs. This is shown in Fig.~\ref{figL}, where the steady state value of $\langle Y^2 \rangle $ is reported versus the mirror distance, with (blue lines) and without (red lines) the effect of CSL.
In the plots, the cavity length $L=R_c$ corresponds to a confocal cavity, whereas the largest value of $L$ approaches the limit of a concentric cavity $L=2R_c$. We remark that the cavity is unstable for larger $L$, so that the presented results cover all the possible geometric configurations of a stable symmetric Fabry-P\'erot resonator.
We observe, in Figs.~\ref{figL} (a) and (b) that as the trapping frequency $\omega$ and the linearized coupling $G$ are gradually reduced, the effect of the collapse-induced diffusion becomes more and more distinguishable. Specifically, the presence of spontaneous collapses is signaled by a change in the slope of these curves for $L>R_c$. When the value of $\lambda$ is decreased, as in Figs.~\ref{figL} (c) and (d), the effect of CSL becomes less and less visible, although here results similar to those reported in Fig.~\ref{testlimit} can be achieved at a slightly larger temperature. In particular, at $\lambda=10^{-12}$ s$^{-1}$, $10^{-11}$ s$^{-1}$ and $10^{-10}$ s$^{-1}$ the relative difference between the two curves, when $L$ is large, is of $\sim$1.5\%, $\sim$12\% and $\sim$30\%, respectively. This implies that both the precision of measurement of the optical phase and the precision with which the experimental parameters are calibrated and kept fixed in the repeated experimental runs must be smaller than the above relative difference. That is, with a parameter calibration and measurement precision of no more than $\sim$1.5\%, one could discriminate the CSL down to $\lambda=10^{-12}$ s$^{-1}$.

\section{conclusions}
\label{concl}

The presented results suggest a general strategy to probe the possible effect of the CSL model and to discriminate it from the effect of other sources of decoherence. It is designed to work with levitated nanospheres in optical Fabry-P\'erot cavities.
The very high mechanical quality factor of these systems makes them
the ideal platforms to test the CMs. Moreover, our approach is effective when the pressure and temperature of background gas are sufficiently small in order to make the corresponding noise negligible. In particular, we have discussed results for the optical phase quadrature $Y$ and we have demonstrated that the presence of the CSL effect can be tested by investigating the stationary behavior of $\langle Y^2\rangle$ as a function of the trapping frequency $\omega$ and of the cavity size $L$. Firstly, we have shown that when $G$ is sufficiently small, the results without CSL are basically independent from $\omega$, while they increase rapidly as $\omega$ decreases when the collapse noise is taken into account, hence providing a distinct signature of the CSL effect. Then we have also shown that by tuning the cavity length and proportionally also the field powers and frequency, the ratios between the optomechanical parameters (frequencies, coupling and photon loss rate) can be kept fixed, hence keeping the system in the same optomechanical regime, while the diffusion rates by light scattering and by CSL exhibit diverging behavior. As a consequence the results for $\langle Y^2\rangle$ reflect a similar behavior, and thus the effect of the CM can be clearly discriminated.

In our analysis we have considered an optomechanical system comprising a diamond nanosphere of radius 100 nm trapped, by an optical dipole trap of a few
kilohertz, inside a Fabry-P\'erot cavity with finesse of $10^5$ and length of a few centimeters. Similar systems have been described in
Refs.~\cite{Barker,Raizen,gieseler,Markus}.
We have thereby demonstrated that these protocols can be employed to test
the strength of $\lambda$ to values as low as $10^{-12}$ s$^{-1}$ with realistic parameters. This value is essentially limited by the temperature and pressure that can be achieved in experiments (we have considered values as low as $P_a=10^{-12}$ Torr and $T=10$ mK, in the results versus $\omega$, and $T=100$ mK, in the results versus $L$) and  lower values of $T$ and $P_a$ would allow to test even lower values of $\lambda$.
These values of temperature and pressure, although challenging, have been already
discussed in various experiments with cold atoms \cite{burrage} and are expected to be achievable in the near future also in experiments involving nanospheres. We have proved that these results can be obtained by the measurement of the phase quadrature of the cavity field, which can be accessed with standard optical techniques. On the other hand, we remark that similar considerations and results are in principle valid for any observable of the optomechanical system. This unique versatility makes the presented proposal very promising for an actual test of the CSL theory.

\section*{ACKNOWLEDGMENT}

J.L. thanks S. Nimmrichter and M. Paternostro for valuable feedback and discussions. This work has been supported by the European Commission (ITN-Marie Curie project cQOM and FET-Open Project iQUOEMS), MIUR (PRIN 2011), National Natural Science Foundation of China (Grant No. 11504218), and Natural Science Foundation of Shanxi (Grant No. 2013021005-2).

\section*{APPENDIX}

In this Appendix we briefly discuss the effects of blackbody radiation in our model. In our treatment we have neglected the effect of blackbody radiation which is very small in the parameter regime addressed in this paper. In general the trapping and cavity light can heat up the particle and the corresponding emitted black-body photons can act as a noise source for the particle motion (absorption of black-body photons is less relevant)~\cite{Chang,nanosphere}. This effect can be estimated using the approach described in the Supporting Information of Ref.~\cite{Chang}. Here we consider diamond which is transparent from ultra-violet to infrared wavelengths, with small absorption due to the quantity and the quality of possible impurities. Using the theory of Ref.~\cite{Chang} with a relatively large absorption coefficient of $1\,$m$^{-1}$ at the cavity field wavelength and a complex relative permittivity with an imaginary part of $10^{-3}$ at the black-body wavelength, we find an internal temperature of the particle of a few hundred degrees Kelvin, and a corresponding diffusion rate that in the worst case is three order of magnitude smaller than the smallest diffusion rate that we have included in our description.

\end{document}